\documentclass[11pt]{article}
\usepackage{pst-all}
\usepackage{pstricks}
\usepackage{pstcol,pst-fill,pst-grad}
\usepackage{amsfonts}
\usepackage{graphicx}
\usepackage{amsmath}

\newcommand{\be}{\begin{equation}}
\newcommand{\ee}{\end{equation}}
\newcommand{\bea}{\begin{eqnarray}}
\newcommand{\eea}{\end{eqnarray}}
\input{tcilatex}
\begin{document}

\title{\textbf{\ Exotic Hadrons and Underlying Z}$_{\mathbf{2,3}}$\textbf{\
Symmetries}}
\author{\textbf{Adil Belhaj}$^{1}$\textbf{\thanks{%
belhaj@unizar.es}},$\;$ \textbf{Salah Eddine Ennadifi$^{2}$\thanks{%
ennadifis@gmail.com},$\;$ Moulay Brahim Sedra$^{3}$\thanks{%
sedra@ictp.it} } \\
{\small $^{1}$D\'{e}partement de Physique, Facult\'{e}
Polydisciplinaire,
Universit\'{e} Sultan Moulay Slimane, B\'{e}ni Mellal, Morocco}\\
{\small $^{2}$LPHE-MS, Facult\'{e} des Sciences, Universit\'{e}
Mohammed V, Rabat, Maroc } \\
{\small $^{3}$ D\'{e}partement de Physique, SIMO-FSK, Facult\'{e}
des Sciences, Universit\'{e} Ibn Tofail K\'{e}nitra, Morocco} }
\maketitle

\begin{abstract}
The recent observation of higher quark combinations, tetraquarks and
pentaquarks, is a strong indication of more exotic hadrons. Using
$Z_{2} $ and $Z_{3}$ symmetries and standard model data, a general
quark combination producing new hadronic states is proposed in terms
of polygon geometries according to  the Dynkin diagrams of
$\widehat{A}_{n}$ affine Lie algebras. It has been shown that
Z$_{\mathbf{2,3}}$ invariance is crucial in the determination of the
mesonic or the baryonic nature of these states. The hexagonal
geometry is considered in some details producing both mesonic and
baryonic states. A general class of this family is also
presented.\newline

\end{abstract}

\newpage

For a long time, it has been realized that the experimental situation in the
Quantum Chromodynamics (QCD) is quite stable and well understood \cite{1,2}.
This describes the interaction between quarks carrying the color charge of
the strong interaction by force mediating gluons which can not only couple
to quarks but also to themselves, due to the $\mbox{SU(3)}_{C}$ gauge
theory. They carry color charge. Although $\mbox{SU(3)}_{C}$ describes the
strong interactions of particles, until today experiments have not been able
to detect colored objects directly. However, only colorless hadrons can be
observed. The explanation of this fact and the combination of hadrons from
quarks and gluons remain incomplete.\newline
Understanding how matter is formed is one of the big problem in modern
particle physics. It is known that the biggest part of the observable matter
is made of strongly interacting quarks and gluons. The standard model (SM)
of particle physics constitutes one of the most succeeded realizations in
this field. Based on the SM symmetry group $\mbox{SU(3)}_{C}\times %
\mbox{SU(2)}_{L}\times \mbox{U(1)}_{Y}$ describing strong, weak and
electromagnetic interactions, it provides an elegant theoretical framework
which is able to describe with a high precision the known experimental
phenomenon in modern physics \cite{3,4}. However, with the progressive
probes of high energies in experiments and results become more precise, the
SM satisfactory is still far and physics beyond is widely expected to reside
with new elementary particles, hadrons, and symmetries \cite{5,6}.\newline
The richness of the hadronic spectrum could be revealed by a fast look into
the particle spectrum \cite{12}. The large number of the hadronic states has
clearly suggested the existence of a deeper level where the messy hadronic
world can be easily understood in terms of a few constituent of spin- $1/2$
quark flavors \cite{7}. At this level, in agreement with the QCD, almost all
observed states could be clearly identified as two- or three-quark states.
More precisely, quark-antiquark $q\overline{q}$ are known as mesons and
three quark states $qqq$ are known as baryons. In this picture, the entire
hadronic spectrum could be nicely classified as the colorless mesons and
baryons. \newline
In the last years, results in the strong sector $\mbox{SU(3)}_{C}$ have
opened new windows in the understanding of flavor physics including quark
configurations \cite{7,8,9,10}. In particular, it has been suggested some
four quark states. This suggestion has been supported by an observation of
new resonances corresponding to four quark states around $2$-$5$ $GeV$. They
belong to a strange series of particles known by $X$, $Y$ and $Z$ states.
They were named $X$, $Y$, $Z$ \cite{NEW1, NEW2, NEW3, NEW4}. Pentaquark
states discovery was also suggested many years ago by several experiments.
However, due to poor data and statistical analysis such claims were not
accepted. Very recently, new results consistent with pentaquark states was
reported by the LHCb collaboration at CERN \cite{NEW5}. Decay experiments
are expected to run soon in order to determine its nature with more
precision. Motivated by these recent works, there is no reason to ignore
higher quark combinations.\newline
In this work, we aim to give a possible generalization dealing with hadronic
states using a geometric approach based on discrete symmetries derived from
the SM gauge group. More precisely, the centers of the SU(3) and SU(2) Lie
symmetries of the SM will guide us to specify the possible orders for the
generalization. To keep contact with the observed physics, we present a
general quark configuration of the mesonic and baryonic states exhibiting $%
Z_{2}$ and $Z_{3}$ global geometrical symmetries. In fact, these symmetries
have been not explored in the related physics. It does not mean that there
have no roles in modern particle physics. Here, we show up their importance.
In particular, they allow one to propose possible higher quark states beyond
the usual hadrons known in the SM.\newline

A close inspection shows that the $Z_{2}$ and $Z_{3}$ symmetries can produce
a higher combination of quarks associated with exotic hadronic states. To do
so, it is convenient to use the following notation,
\begin{equation}
M:|q\overline{q}\rangle ,\qquad B:|qqq\rangle .  \label{eq1}
\end{equation}%
It is observed that these states are invariant under $Z_{2}$ and $Z_{3}$
symmetries respectively. The mesonic states containing $q$ and $\overline{q}$
are invariant under the $Z_{2}$ symmetry acting as,
\begin{equation}
q\rightarrow -q,\qquad \overline{q}\rightarrow -\overline{q}.  \label{eq2}
\end{equation}%
However, the baryonic states are invariant under the $Z_{3}$ symmetry acting
as
\begin{equation}
q\rightarrow wq,\quad w^{3}=1.  \label{eq3}
\end{equation}%
It is important to recall that the quadratic combination $|qq\rangle $ is
invariant under $Z_{2}$ acting only as $q\rightarrow -q$. However, it is not
allowed by the confinement phenomenon\footnote{%
Only colorless hadrons can be observed.}. This kind of states and their
generalization should be projected out. \newline
According to the SM data and the $Z_{2,3}$ symmetries, we give an extended
combination for the allowed hadronic states. To be precise, we explore
techniques of polygon geometries to support the existence of such exotic
states. Our analysis will be based on geometric shapes which are invariant
under the above mentioned symmetries. In fact, each hadronic state will
correspond to an allowed polygon geometry playing similar role as graph
theory used to encode physical data including gauge and matter fields
derived from string theory compactification using quiver method \cite{110}.
In this scenario, the $Z_{2}$ and $Z_{3}$ symmetries will be crucial in the
determination of the nature of such states. To show how the philosophy works
in practice, let us first start by giving the two leading polygon geometry
pieces playing a primordial role in the building block of higher hadronic
states. In this way, the mesonic state (\ref{eq1}) is represented by two
connected nodes, and the leading baryonic state $|qqq\rangle $ is
represented by three nodes forming a triangle. At first sight, it seems that
these graphs share similarities with the toric geometry realization of
complex manifolds\cite{a1}. However, the SM Lie symmetries push us to think
about polygon geometries associated with such symmetries. Considering closed
quiver diagrams and taking into account of $Z_{2,3}$ symmetries, the su(2)
and su(3) Lie symmetries should be shown up in the discussion. An
investigation reveals that these quivers can be identified with the
corresponding affine Dynkin geometries \cite{a2}. This is illustrated in
figure 1.
\begin{figure}[h]
\begin{center}
{\includegraphics[scale=0.5]{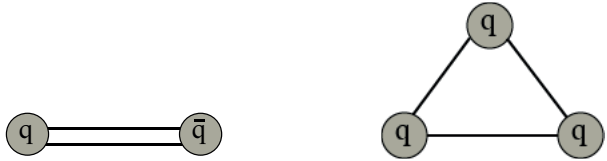}}
\end{center}
\caption{Standard hadrons: mesons (left) and baryons (right).}
\end{figure}
Inspired by the recent theoretical and experimental works \cite{9,10,11},
the extended hadronic states will be represented by closed Dynkin diagrams
associated with the Affine $\widehat{A}_{n}$ series. This geometric approach
based on the $Z_{2}$ and $Z_{3}$ symmetries is interesting since one can
control completely the engineering of allowed diagrams associated with
possible hadronic states.\newline
Naturally, the first extended polygon geometry will be associated with four
nodes corresponding to the $\widehat{A}_{3}$ affine Dynkin diagram. A
priori, there are many $q$ and $\overline{q}$ systems which can be placed on
these four nodes. However, the above physical requirement allow only one
possible configuration invariant under a geometric $Z_{2}$ leading to a
mesonic state $|q\overline{q}q\overline{q}\rangle $. Such states have been
investigated in \cite{9,10}, and have been supported by the recent
experimental result \cite{11}. Roughly, the corresponding diagram and the
geometric $Z_{2}$ symmetry are represented in the figure 2.
\begin{figure}[h]
\begin{center}
{\includegraphics[scale=0.5]{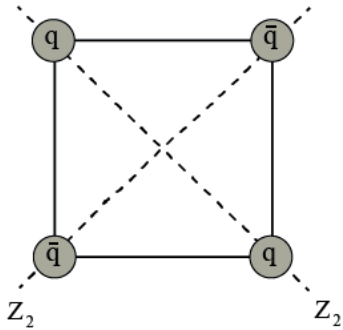}}
\end{center}
\par
\vspace*{-.2cm}
\caption{Tetraquark mesons.}
\end{figure}
With such four quarks constituents, several ways of combinations are
possible. Indeed, a tetraquark can either contain four quarks $|q\overline{q}%
q\overline{q}\rangle $ or something resembling two ordinary mesons bound
together $\left \vert \left( q\overline{q}\right),\left( q\overline{%
q}\right) \right \rangle $ (a dimeson). Reported examples of such
tetraquark states are the mesons $X(3872)$, $Y(4140)$,  $Zc(3900)$
and $Z\left(
4430\right) ^{-}$. They differ by theirs  flavors content $\left \vert q%
\overline{q}q\overline{q}\right \rangle _{X}\neq |q\overline{q}q\overline{q}%
\rangle _{Y}\neq |q\overline{q}q\overline{q}\rangle _{Z_{C}}\neq |q\overline{%
q}q\overline{q}\rangle _{Z}$ which are still being under investigated to
characterize their determined masses and quantum numbers. Concretely, the
heaviest one, $Z\left( 4430\right) ^{-}$, with a negative charge -1 and a
mass around $4,430$\textbf{\ }$GeV$, has been definitely identified with the
tetraquark meson $|q\overline{q}q\overline{q}\rangle \equiv \left \vert c%
\overline{c}d\overline{u}\right \rangle _{Z}$\textbf{\ }\cite{NEW4}. It is
worth noting that the $|qqqq\rangle $, for instance, is not allowed as
required by the confinement phenomenon.\newline
The next possible five quarks state $|q\overline{q}qqq\rangle $ has been
very recently also identified from the decay of the bottom lambda baryon $%
\Lambda _{b}^{0}$ into a $J/\Psi (c\overline{c})$ meson, a kaon $K^{-}\left(
s\overline{u}\right) ${\ and a proton }$p\left( uud\right) $ by the LHCb
with two hidden-charm pentaquarks as $|q\overline{q}qqq\rangle \equiv |uud%
\overline{c}c\rangle _{P_{C}^{+}}$ being a charmonium-pentaquarkn \cite{NEW5}%
. Such state can be combined into two possible combinations: a five quark
bag $|q\overline{q}qqq\rangle $ an ordinary meson-baryon bounded together $%
|\left( qqq\right) ,\left( q\overline{q}\right) \rangle $. It is interesting
to note that, with respect to the $Z_{2,3}$ symmetries, the latter
combination is more likely. \newline
After a graphic examination based on the $Z_{2,3}$ invariance, the next
fantastic example concerns six nodes associated with the hexagonal geometry.
The appearance of this geometry is the more remarkable one since it appears
naturally in many places in physics including even in the SU(3) gauge theory
describing the strong interaction \cite{a3}. Indeed, the gluon gauge fields
are associated with the root systems of the corresponding su(3) finite Lie
symmetry. Its Weyl group is the symmetric group $S_{3}=D_{3}$ of order 6
which acts transitively and freely on the hexagon. It is obvious to see that
this group contains $Z_{2}$ and $Z_{3}$ as subgroups. At this level, we
remark that the presence of such symmetries seem to be useful in the present
discussion. We believe that this feature would play an important role in the
understanding of the hexagonal hadrons. In fact, we will see that this
geometry produces both mesonic and baryonic states depending on the chosen
symmetry. Later on, we show that this example constitutes in fact particular
geometries of infinitely many possible quark configurations giving rise to
both mesonic and baryonic states. For the hexagonal structure, we can
obviously determine the nature of hadronic state by taking only one discrete
symmetry. It is observed that the $Z_{3}$ geometric invariance generates the
baryonic state $|qqqqqq\rangle $. Its complex conjugate is also allowed
giving the sate $|\overline{q}\overline{q}\overline{q}\overline{q}\overline{q%
}\overline{q}\rangle $. However, the $Z_{2}$ geometric invariance gives only
a mesonic state denoted by $|q\overline{q}q\overline{q}q\overline{q}\rangle $
in accordance with the known data. The corresponding diagrams are
represented in the figure 3.

\begin{figure}[h]
\begin{center}
{\includegraphics[scale=0.5]{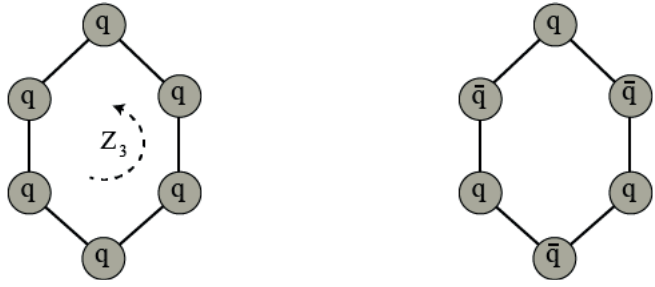}}
\end{center}
\par
\vspace*{-.2cm}
\caption{Hexaquark Baryons (left) and hexaquark mesons (right).}
\end{figure}
As for the four state quark constituents, such six quark ones
involves several ways of combinations are possible. Indeed, a
hexaquark can either contain six quarks $|qqqqqq\rangle $ or
something resembling two ordinary baryons bound together $\left
\vert \left( qqq\right) \mathbf{,}\left( qqq\right) \right \rangle $
a dibaryon, or three quarks and three antiquarks
$\left \vert \left( qqq\right) \mathbf{,}\left( \overline{q}\overline{q}%
\overline{q}\right) \right \rangle $ a baryon antibaryon or six antiquarks $%
\left \vert \left( \overline{q}\overline{q}\overline{q}\right) \mathbf{,}%
\left( \overline{q}\overline{q}\overline{q}\right) \right \rangle $
(a diantibaryon). Once formed, dibaryons are predicted to be fairly
stable, i.e, the diproton and dineutron. The existence of a stable
dibaryon having the light quark composition $|qqqqqq\rangle =\left
\vert udsuds\right \rangle $ is possible. To detect such states,
many  experiments have been suggested for their decays and
interactions where several candidates were observed but they are not
yet confirmed \cite{NEW6}. The last potential one,
called $d^{\mathbf{\ast }}(2380)$, was detected at about $2,380$ $GeV$%
\textbf{\ }for\textbf{\ }$10^{\mathbf{-23}}s$ \cite{NEW7}.\newline
We can now generalize to higher geometries involving more than six nodes. To
get a mesonic states, the geometry should consist of $2+2n$ nodes according
to the $\widehat{A}_{1+2n}$ Dynkin diagram. SM and the $Z_{2}$ invariance
require a system of $(1+n)$ quarks and $(1+n)$ antiquarks. However, the
geometry corresponding to a baryonic state contains a collection of $3+3m$
nodes according to the $\widehat{A}_{2+3m}$ affine Dynkin diagram. The
corresponding states will be denoted by $|q\overline{q}\dots q\overline{q}%
\ldots q\overline{q}\rangle $ and $|q\dots q\ldots q\rangle $ respectively
and are illustrated by the diagrams presented in figure 4.

\begin{figure}[h]
\begin{center}
{\includegraphics[scale=0.5]{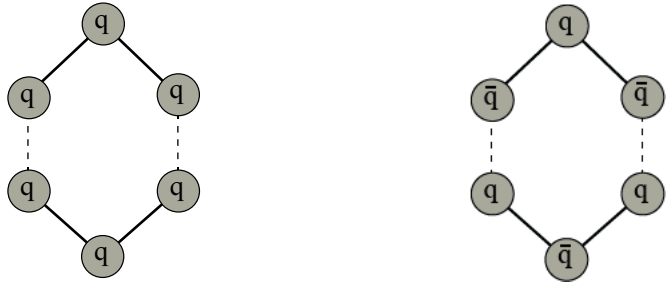}}
\end{center}
\par
\vspace*{-.2cm}
\caption{Higher hadrons: higher baryons (left) and higher mesons (right).}
\end{figure}
It is worth noting to comment this generalization involving two interesting
features. First, it is noted that the case $n=m=0$ corresponds to the
standard baryonic and mesonic states. Second, it is observed that both the
baryonic and mesonic states can appear when $m=2j+1$ and $n=2m-j$ where $j$
is a positive integer. This family of geometry goes beyond the case of the
hexagon associated with $j=0$.\newline

In this work, we have proposed new quark combinations generating exotic
hadronic states. The theoretical analysis has been relied on polygon
geometries sharing similarities with the $\widehat{A}_{n}$ affine Dynkin
diagrams. Inspired by the QCD description of the strong interaction and its
effective treatment of the quark states arising in standard pictures of the
mesonic and baryonic states, we have explored the $Z_{2,3}$ geometrical
symmetries to construct possible higher quark states going beyond the
standard hadrons. These symmetries have played an essential role for
identifying the mesonic or the baryonic nature of the corresponding states.
Although, we have tried to give a theoretical argument behind the existence
of higher quark combinations, this present study deserves a deeper
investigation since the observation of more exotic hadrons remains enhanced.

{\bf Acknowledgments}:   AB would like  to thank his family in
T\'emara (Morocco) for  hospitality,  where  a part of the research
reported in this work was conducted.  He would like  to thank also
his mother Haja Fatima for all kind of supports.

\end{document}